\documentclass{svmult}

\usepackage{mathptmx}       
\usepackage{helvet}         
\usepackage{courier}        
\usepackage{type1cm}        
\usepackage{makeidx}         
\usepackage{graphicx}        
\usepackage{multicol}        
\usepackage[bottom]{footmisc}

\usepackage{textcomp}        
\usepackage{amssymb}        

\makeindex             

\begin{document}

\title*{Atomistic simulations of electrolyte solutions and hydrogels with explicit solvent models}
\titlerunning{Atomistic simulations of electrolyte solutions and hydrogels}
\author{Jonathan Walter \inst{1} \and Stephan Deublein \inst{1} \and Steffen Reiser \inst{1} \and Martin Horsch \inst{1} \and Jadran Vrabec \inst{2} \and Hans Hasse \inst{1}}
\authorrunning{J. Walter, S. Deublein, S. Reiser, M. Horsch, J. Vrabec and H. Hasse}
\institute{Lehrstuhl f\"ur Thermodynamik, Technische Universit\"at Kaiserslautern, Erwin-Schr\"odinger-Stra{\ss}e 44, 67663 Kaiserslautern, Germany \texttt{martin.horsch@mv.uni-kl.de} \and Lehrstuhl f\"ur Thermodynamik und Energietechnik, Universit\"at Paderborn, Warburger Stra{\ss}e 100, 33098 Paderborn, Germany}
%
%
\maketitle

\section{Introduction}
\label{sec:1}

Two of the most challenging tasks in molecular simulation consist in capturing the properties of systems with long-range interactions (e.g. electrolyte solutions), and of systems containing large molecules such as hydrogels. These tasks become particularly demanding when explicit solvent models are used. Therefore, massively parallel supercomputers are needed for both tasks.\\
For the development and optimization of molecular force fields and models, a large number of simulation runs have to be evaluated to obtain the sensitivity of thermodynamic properties with respect to the model parameters. This requires both an efficietn work flow and, obviously, even more computational resources. The present work discusses the force field development for electrolytes regarding thermodynamic properties of their solutions. Furthermore, simulation results for the volume transition of hydrogels in solution containing electrolytes are presented. Both applications are of interest for engineering. It is shown that the properties of these complex systems can be reasonably predicted by molecular simulation.

\section{Development of force fields for alkali and halogen ions in aqueous solution}
\label{stephan_1}

\subsection{Outline}

The simulation of electrolyte systems is computationally very expensive due to the long-range interactions, which have to be taken into account by suitable algorithms. Examples of such algorithms are the classical Ewald summation~\cite{Ewald1921} and its derivatives like particle mesh Ewald summation~\cite{Darden1993} or the particle particle/particle mesh method~\cite{Eastwood1980}.\\
Early efforts in this research area were mainly directed to the development of ion force fields, which were capable of reproducing static structural properties of solutions~\cite{Rao1990,Fumi1964,Dang2002}. Recently, however, these models have been proven to be too inaccurate for the prediction of basic thermodynamic properties~\cite{Hess2006,Walter2010}. Since the models were parameterized solely at short distances, long distance effects are underestimated. These effects can be seen in particular in the density and the activity of electrolyte solutions. \\
In the present work new atomistic models for alkali and halogen ions are presented, which accurately describe not only structural properties of aqueous electrolyte solutions but also basic thermodynamic properties like their density. The main focus of the model development is placed on transferability, i.e. the ion models are intended to describe solution properties independently of the cation/anion combination. \\
The ions were modeled as Lennard-Jones (LJ) spheres with superimposed charges of \textpm1 in units of elementary charges, located in the center of mass. For the solvent water, the SPC/E~model~\cite{Berendsen1987} was used, which consists of one LJ site and three point charges. \\

\subsection{Simulation details}
\label{stephan_2}
All simulations in Section \ref{stephan_1} were performed using the Monte-Carlo technique in the isobaric-isothermal ($NpT$) ensemble at 20~$\ensuremath{{^\circ}}$C and 0.1~MPa. The simulation volume contained a total of 1000~molecules and ions, respectively. The electrostatic long-range interactions were calculated using the Ewald summation with an Ewald parameter $\kappa$ of 5.6/$L$, where $L$ denotes the length of the cubic simulation volume. The dispersive and repulsive long-range contributions were approximated using the assumption of an homogeneous fluid beyond the cut-off distance of~11.9~\AA. The simulation program employed was an extended version of $ms2$~\cite{deublein2011}.\\

\subsection{Solvent model and reduced properties}
\label{stephan_3}
\begin{figure}
\begin{center}
 \includegraphics[width=.75\textwidth]{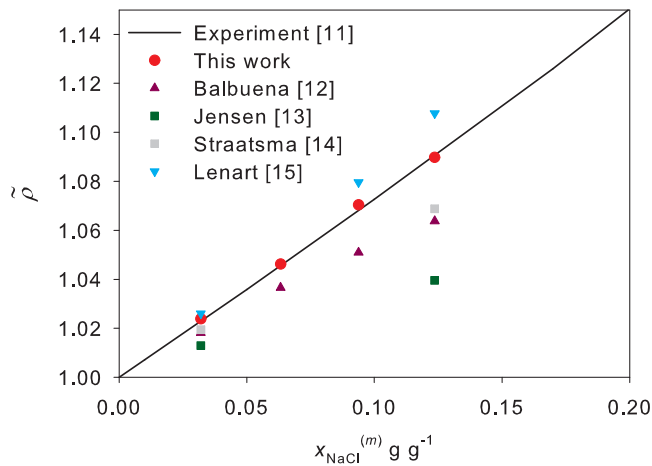}
\caption{Reduced density $\widetilde{\rho}$ as a function of sodium chloride mass fraction. The line indicates the experimental values~\cite{Weast1987}, while the symbols represent simulation results using varying ion force fields~\cite{Balbuena1996,Jensen2006,Straatsma1988,Lenart2007}.}
\label{stephan_fig1}
\end{center}
\end{figure}

For the calculation of aqueous electrolyte solutions, the model for water is crucial, since it represents the largest fraction in the mixture. The SPC/E~\cite{Berendsen1987} water model was employed, which reproduces the density and other thermodynamic properties at chosen conditions in good agreement with experimental data ($\rho_{\mathrm {SPC/E}} (T=20^{\circ} \mathrm{C}, p=0.1 \mathrm{MPa})$~=~999.5~g/l). To minimize the influence of errors in the solvent model on the parameterization of the ion force fields, the reduced density~$\widetilde{\rho}$ of the aqueous solution was chosen as objective function, which is defined as the fraction of the density of the electrolyte solution $\rho_{\mathrm {ES}}$ and the density of the pure solvent $\rho_{\mathrm S}$ at the same temperature and pressure
\begin{equation}
\label{stephan_eq1}
 \widetilde{\rho} = \frac{\rho_{\mathrm {ES}}}{\rho_{\mathrm S}}   \mbox{   .}
\end{equation}
Figure~\ref{stephan_fig1} shows a typical plot of the reduced density of an aqueous electrolyte solution as a function of salt mass fraction using sodium chloride as an example. The dependence of $\widetilde{\rho}$ with $x^{\mathrm m}$ is almost linear. It can also be seen from Figure~\ref{stephan_fig1} that electrolyte force fields from the literature predict this linear correlation, but with a wrong slope. Note also that the solvent activity (or activity coefficient) can be regarded as a normalized property. However, a similar approach fails for properties like the ion self-diffusion coefficient.

\subsection{Ion force field}
\label{stephan_ionparam}
A global parameterization strategy was used in the present study. Due to the simple model for the ions, the overall parameter space is small. It contains the two LJ parameters $\sigma$ and $\varepsilon$ for each ion. The position and magnitude of the charges are constant. In a preliminary analysis, the sensitivity of the reduced density on the Lennard-Jones parameters was evaluated. It shows that the influence of the LJ energy parameter $\varepsilon$ on~$\widetilde{\rho}$ is very small. Therefore, the $\sigma$ parameters were determined by a fit to the reduced density, while the $\varepsilon$ parameters were set here to a constant value, i.e. 200~K for sodium and chloride, based on results of a study on the water activity in the electrolyte solution, which is not described in detail here. \\
The objective function of the optimization used to determine the LJ~$\sigma$ parameters is the slope of the reduced density~$\widetilde{\rho} $ with increasing salt mass fraction at $x^{\mathrm {(m)}} \rightarrow 0$ 
\begin{equation}
\left( \frac{\mathrm{d} \widetilde{\rho}} {\mathrm{d} x^{(m)}} \right)_{x^{\mathrm {(m)}} \rightarrow 0}  = \left( \frac{\mathrm{d} \widetilde{\rho}} {\mathrm{d} x^{(m)}}     (\sigma_{\mathrm {anion}}, \sigma_{\mathrm {cation}} ) \right)_{x^{\mathrm {(m)}} \rightarrow 0}   \mbox{   .}
\end{equation}
In molecular simulations, these slopes were systematically derived for varying values of $\sigma_{\mathrm {anion}}$ and $\sigma_{\mathrm {cation}}$. The data were smoothed by a two dimensional polynomial fit. The ion parameters for all alkali and halogen ions were chosen such that the sum of the squared deviations between the functional fit and the experimental data is minimized.

\subsection{Results}
\label{stephan_results}

\begin{figure}
\begin{center}
\includegraphics[width=.75\textwidth]{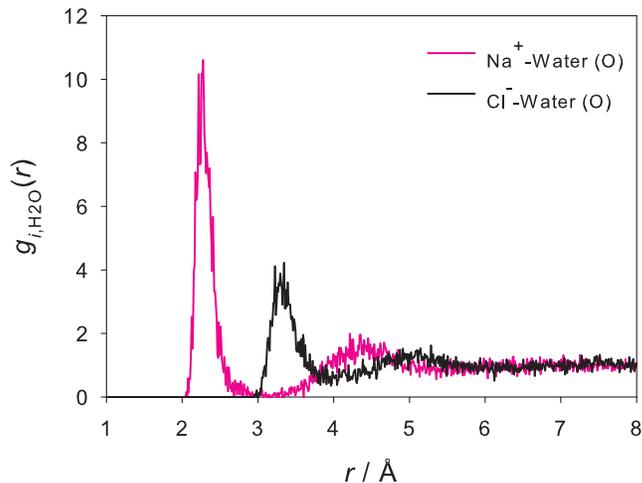}
\caption{Radial distribution function $g_{i\mathrm{,H_2O}}(r)$ of the solvent water around the sodium cation and chloride anion, respectively, at $T=$20~$^{\circ}$C and $p=$0.1~MPa. }
\label{stephan_fig3}
\end{center}
\end{figure}

New atomistic models for alkali and halogen ions were determined that describe the reduced density of the aqueous electrolyte solution in good agreement with experiments. As an example the force field for sodium chloride is shown in Table~\ref{stephan_table_1}. Figure~\ref{stephan_fig1} shows the excellent agreement between the simulation using the new model and the experimental results.
\begin{table}
\caption{Force field of sodium chloride}
\label{stephan_table_1}
\begin{center}
 \begin{tabular}{p{1.7cm}  p{1.5cm} p{1.5cm} c}
\noalign{\smallskip}\hline\noalign{\smallskip}  Ion     &  $\sigma $ / \AA & $\varepsilon$ / K & \\
\noalign{\smallskip}\hline\noalign{\smallskip}   Na$^+$ &       1.88          &       200       & \\
\noalign{\smallskip}\noalign{\smallskip}   Cl$^-$ &       4.41          &       200       & \\
  \hline
 \end{tabular}
\end{center}
\end{table}\\
The new atomistic models for alkali and halogen ions were investigated regarding the representation of structural properties of aqueous solution. Radial distribution functions $g_{ij}(r)$ were used for the characterization which are defined by
\begin{equation}
 g_{ij}(r) = \frac{\rho_{j}(r)}{\rho_{j, \mathrm {bulk}}}   \mbox{   .}
\end{equation}
Here $\rho_j(r)$ is the density of component $j$ as a function of the distance $r$ between two ions of component $i$ and $j$, respectively, and $\rho_{j, \mathrm{bulk}}$ is the number density of component $j$ in the bulk phase. For the characterization of aqueous electrolyte solutions, the radial distribution function of water around the ions is of particular interest. In this case, water is represented by the position of the oxygen atom. For a solution of sodium chloride in water, $g_{i,\mathrm{H_2O}}(r)$ is shown in Figure~\ref{stephan_fig3}.
\\
The locations of the first maximum and minimum, respectively, for both ions are in good agreement with experimental measurements~\cite{Marcus1988}, as shown in Table~\ref{stephan_table_2}. The hydration shell of the cation is more attached to the ion than of the anion, which is represented by the height of the first peak in the distribution function.
\begin{table}
\caption{Location of the first maximum $r_{\mathrm {max}}$ and minimum $r_{\mathrm {min}}$ of the radial distribution function $g_{i,\mathrm{H_2O}}(r)$ for an aqueous sodium chloride solution. The simulative results are compared to experimental data from the literature~\cite{Marcus1988}.}
\label{stephan_table_2}
\begin{center}
 \begin{tabular}{p{1.7cm}   p{1.5cm}  p{1.5cm}  p{1.5cm}  p{1.5cm} c}
\noalign{\smallskip}\hline\noalign{\smallskip}      Ion $i$  & $r^{\mathrm {Sim}}_{\mathrm {max}}$ / \AA & $r^{\mathrm {Exp}}_{\mathrm {max}}$  / \AA & $r^{\mathrm {Sim}}_{\mathrm {min}}$  / \AA & $r^{\mathrm {Exp}}_{\mathrm {min}}$  / \AA  &\\
\noalign{\smallskip}\hline\noalign{\smallskip}     Na$^+$ &       2.2          &      2.3          &   3.0          &       3.0       &\\
\noalign{\smallskip}\noalign{\smallskip}     Cl$^-$ &       3.4          &      3.3          &   3.9          &       4.0       &\\
  \hline
 \end{tabular}
\end{center}
\end{table}\\
The attraction of the solvent to the cation is also visible in the hydration number around the ions, which describes the number of solvent molecules in the closest vicinity to the solute $i$ and is defined by
\begin{equation}
 n_{i,\mathrm{H_2O}} = 4\pi \rho_{\mathrm{H_2O}} \int_0^{r_{\mathrm{min}}} r^2g_{i,\mathrm{H_2O}}(r) \mathrm{d}r \mbox{   . }
\end{equation}
Here, $\rho_i$ defines the number density of component $i$ and $r_{\mathrm{min}}$ the distance up to which the hydration number is calculated. For the first shell, the value of $r_{\mathrm{min}}$ is chosen to be the distance of the first minimum in the radial distribution function. For sodium chloride for example, the calculation of hydration numbers reproduces experimental results in good agreement, cf. Table~\ref{stephan_table_3}.
\begin{table}
\caption{Hydration number $n$ of Na$^+$ and Cl$^-$ in aqueous solution. The simulative results are compared to experimental data from the literature~\cite{Marcus1988}.}
\label{stephan_table_3}
\begin{center}
 \begin{tabular}{p{1.7cm}  p{1.5cm} p{1.5cm} c}
\noalign{\smallskip}\hline\noalign{\smallskip}        Ion    &  $n^{\mathrm {Sim}}$ & $n^{\mathrm {Exp}}$& \\
\noalign{\smallskip}\hline\noalign{\smallskip}        Na$^+$ &       5.6          &       5 - 6      & \\
\noalign{\smallskip}\noalign{\smallskip}        Cl$^-$ &       7.5          &       7 - 8      & \\
  \hline
 \end{tabular}
\end{center}
\end{table}

\subsection{Computational demands}
\label{stephan_compdemands}

Typical simulations to generate data points for the studies discussed in Section~\ref{stephan_1} were carried out on 16~CPUs running for 72~hours. For the prediction of other thermodynamic data like activity coefficients, up to 32 CPUs running for 72~hours depending on the system are required. For these simulations a virtual ram of 216~MB was required.

\section{Self-diffusion coefficients of solutes in electrolyte systems}
\label{sec:3}

\subsection{Outline}
The self-diffusion coefficient is, in comparison to the density, an individual property of the different solute species and 
the solvent in electrolyte systems. Self-diffusion coefficients of anions and cations in aqueous solution are experimentally accessible and numerous experimental data are available in the literature \cite{dna_wang,dcl_nielsen}. Therefore, it is worthwhile trying to fit model parameters of ion force fields to self-diffusion coefficient data. As a suitable strategy for determining the size parameters $\sigma$ is available (cf. Section \ref{stephan_1}), the question may be raised whether self-diffusion coefficient data are useful for determining the energy parameters $\epsilon$.\\
In molecular simulations, the self-diffusion coefficient is usually determined by time and memory consuming methods, which require simulations of large systems. Examples for such methods in equilibrium molecular dynamics are the mean square displacement \cite{mean_square} and the Green-Kubo formalism \cite{green,kubo}. In this formalism, the self-diffusion coefficient is related to the time integral of the velocity auto-correlation function. The calculation of self-diffusion coefficients of solutes in electrolyte systems are computationally expensive due to additional time consuming algorithms (e.g. Ewald summation \cite{Ewald1921}) that allow for a truncation of ionic interactions in molecular simulations.\\
In aqueous solution, the cations and anions are surrounded by a shell of strongly bonded water molecules (hydration shell).
These hydrated ions diffuse within a bulk fluid, which is itself also highly structured. Therefore, the mobility and accordingly the self-diffusion coefficient of ions is strongly related to the structure of the water molecules around the ions \cite{structure_koneshan}.

\subsection{Methods}
The investigated solution consisted of sodium and chloride ions as solutes and explicit water as solvent. The ions were modeled as Lennard-Jones spheres with a central charge. Size parameters $\sigma$ for different ion force fields were determined as discussed in Section \ref{stephan_1}. The energy parameter $\epsilon$ was modified in a range of 50 to 250 K. Water models were taken from the literature.\\
First, the density of the solution was determined in an isobaric-isothermal (\textit{NpT}) molecular dynamics (MD) simulation at a desired temperature and pressure. Then, the velocity auto-correlation function and, according to the Green-Kubo formalism \cite{green,kubo}, the self-diffusion coefficients were determined in an isochoric-isothermal (\textit{NVT}) MD simulation at this temperature and density. The sampling length of the velocity auto-correlation function was set to 11 ps and the time span between the origin of two auto-correlation functions was 0.06 ps. The separation between the time origins was chosen such that all autocorrelation functions have decayed at least to 1/\textit{e} of their normalized value. For both \textit{NpT} and \textit{NVT} simulations, the molecular dynamics unit cell with periodic boundary condition included 4420 water molecules, 40 sodium and 40 chloride ions. The long-range charge-charge interactions were calculated using Ewald summation \cite{Ewald1921}. The simulation program employed was an extended version of \textit{ms2} \cite{deublein2011}, which is developed by our group.

\subsection{Results}
The self-diffusion coefficients of anions and cations determined in molecular simulation largely depend on the used molecular model of water, as the mobility of the ions is influenced by the hydration shell and the structure of the surrounding water.

\subsubsection*{Self-diffusion coefficient of water models}
The accuracy of the estimated self-diffusion coefficients of pure water for different water models is verified with respect to experimental data \cite{waterexp}. The self-diffusion coefficients were determined with the Green-Kubo formalism \cite{green,kubo} in molecular dynamics simulations at a temperature of 25 \ensuremath{{^\circ}}C and a pressure of 0.1 MPa. For this study, three commonly used rigid nonpolarizable molecular water models of united-atom type, namely SPC/E \cite{Berendsen1987}, TIP4P \cite{tip4p} and TIP4P/2005 \cite{tip4p2005}, were chosen.
\begin{table}
\caption{Self-diffusion coefficients of pure water at 25 \ensuremath{{^\circ}}C and 0.1 MPa of different molecular water models. The number in parenthesis indicates the uncertanty of the last digit.}
   \centering
   \begin{tabular}{p{1,7cm} p{1,5cm} p{1,5cm} p{1,5cm} p{1,5cm}}\noalign{\smallskip}\hline\noalign{\smallskip}
   Model     & SPC/E& TIP4P& TIP4P/2005& Experiment\\ \noalign{\smallskip}\hline\noalign{\smallskip}
   $D_{\textrm{w}}$  [$\textrm{m}^2 \textrm{s}^{-1}$] & 26.2 (1)& 36.7 (2)& 21.9 (1)& 22.3 (-) \\ \noalign{\smallskip}\hline
    \end{tabular}
\label{waterdiffcoeff}
\end{table}\\
The determined self-diffusion coefficients of the different water models vary over a wide range, cf. Table \ref{waterdiffcoeff}. For TIP4P/2005, the self-diffusion coefficient of pure water is in good agreement with the experimental value. In contrast, both the SPC/E and the TIP4P model overestimate the mobility of water molecules in pure water. The obtained values for the self-diffusion coefficient are in good agreement with the results published by other authors \cite{waterguevara}.

\subsubsection*{Model development}
\begin{figure}
  \centering
  \includegraphics[width=0.75\textwidth]{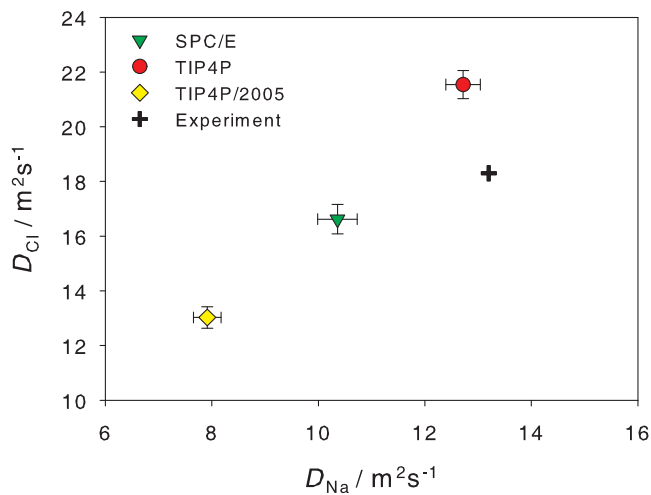}
  \caption{Self-diffusion coefficient of sodium ($D_{\textrm{Na}}$) and chloride ($D_{\textrm{Cl}}$) in aqueous solution at 25 \ensuremath{^\circ}C
  and 0.1 MPa for different water models. The energy parameter $\epsilon$ for both sodium and chloride force fields was set to 200 K. Experimental data for sodium \cite{dna_wang} and chloride \cite{dcl_nielsen} were taken from the literature.}
  \label{iondiffcoeff}
\end{figure}

For fitting the energy parameter $\epsilon$ of the ion force fields to experimental self-diffusion coefficients of anions and cations in aqueous solution, the influence of $\epsilon$ on the determined self-diffusion coefficients in molecular simulations was investigated. In this study, the above mentioned three water models were used.\\
However, it turned out that the number for the energetic parameter $\epsilon$ has no significant influence on the self-diffusion coefficient. This study is not discussed here in detail. In the following, $\epsilon$ = 200 K is used. The results for both sodium and chloride are shown in Figure \ref{iondiffcoeff}.
\\
As can be seen in Table \ref{waterdiffcoeff} and Figure \ref{iondiffcoeff}, there is no correlation between the accuracy of the determined self-diffusion coefficient of pure water and the accuracy of the estimation of ion mobility for the different water models. For example, the TIP4P model significantly overestimates the self-diffusion coefficient of pure water, whereas the simulation results for the self-diffusion coefficients of the sodium and the chloride ion are in fair agreement with experimental data (deviation for sodium of 4\%; deviation for chloride of 18\%).\\
All water models overestimate the interaction between the water molecules in the hydration shell, the sodium ion and the bulk fluid. Hence, the cation mobility is too low. The same is true for the chloride ion only for SPC/E and TIP4P/2005 water models, whereas the TIP4P underestimates the interaction in the hydration shell.

\subsection{Computational demands}
All molecular simulations in Section \ref{sec:3} were carried out with the MPI based molecular simulation program \textit{ms2}, which is developed in our group. The total computing time for determining self-diffusion coefficients of ions in electrolyte systems was 216 hours on 36 CPUs (72 hours for the \textit{NpT} run and 144 hours for the \textit{NVT} run). These simulations require large systems as the accuracy of the Green-Kubo formalism for ions increases with increasing number of solutes and, at the same time, an infinite dilution is aspired. For these simulation a maximum virtual memory of 1.76 GB was used.

\section{Hydrogels in electrolyte solutions}
\label{sec:4}

\subsection{Outline}

Hydrogels are three-dimensional hydrophilic polymer networks. Their most characteristic property is their swelling in aqueous solutions by absorbing the solvent, which is influenced by various factors. Hydrogels can be used in many applications like e.g. superabsorbers such as in diapers \cite{Rehim_2005} and contact lenses \cite{Pavlyuchenko_2009}. To fully exploit the potential of hydrogels in all these applications, it is important to understand, describe and predict their swelling behavior.
The hydrogel which is studied in the present work is built up of poly(N-isopropylacrylamide) (PNIPAAm) cross-linked with N,N'-methylenebisacrylamide (MBA). PNIPAAm is one of the most extensively studied hydrogels in the scientific literature and is mainly used in bioengineering applications \cite{Rzaev_2007}. The degree of swelling in equilibrium of PNIPAAm is significantly influenced by many factors \cite{Huether_2004_2,Huether_2006,Mukae_1993,Crowther_1998}. On the one hand, the swelling depends on the structure of the hydrogel itself, like the type of the monomer, but also the amount and type of cross-linker and of co-monomers. On the other hand, the environment conditions like temperature, type of solvent, solvent composition, electrolyte concentration or pH-value of the solvent influence the swelling. Varying these factors, the hydrogel typically shows a region where it is swollen and a region where it is collapsed. In between those two regions lies the region of volume transition. The solvent composition which is characteristic for that transition is called $\Theta$-solvent here. The $\Theta$-solvent mainly depends on the environmental factors and the nature of the polymer chain but not on the amount of cross-linker.\\
For the quantitative description of the swelling of hydrogels, various types of models are used \cite{Wu_2004}. It is normally not possible to quantitatively predict the swelling of hydrogels or its dependence on factors the models were not adjusted to. With molecular simulation it is possible to predict the swelling of different hydrogels upon varying any environmental factor, as was shown in a previous study \cite{Walter_2010_2}.\\
In the present work, the swelling of PNIPAAm hydrogel is studied with atomistic molecular dynamics simulation. The results for the $\Theta$-solvent are compared to experimental data of PNIPAAm hydrogel as a function of the temperature in water \cite{Zhang_2005}. The experimental data shows that the $\Theta$-solvent of PNIPAAm in different electrolyte solutions follows the Hofmeister series. For this study the electrolyte sodium chloride (NaCl) was considered.

\subsection{Models}

For the molecular dynamics simulations of PNIPAAm in aqueous solutions, the OPLS-AA force field \cite{Jorgensen_1988,Jorgensen_1996} was employed to describe PNIPAAm. It was combined with the SPC/E water model \cite{Berendsen1987}. In previous studies, it was shown that this combination allows predicting the volume transition of PNIPAAm in water as function of the temperature \cite{Walter_2010_2}. Different NaCl models from the literature were used and compared to the NaCl model developed in this work. The used electrolyte models from the literature are GROMOS-96 53A6 (G53A6) \cite{Oostenbrink_2004_2} and KBFF \cite{Weerasinghe_2003}.\\

\subsection{Simulation details}

Molecular simulations of PNIPAAm single chains were carried out with version 4.0.5 of the GROMACS simulation package \cite{Spoel_2005,Hess_2008}. Simulations with PNIPAAm in aqueous NaCl solutions at 25~\ensuremath{{^\circ}}C were performed in order to find the best NaCl model for the simulation of the volume transitions of PNIPAAm in NaCl solutions.\\
In a previous study, it was shown that the amount of cross-linker of the hydrogel has no significant influence on the $\Theta$-solvent of the hydrogel and that this value is the same as for the PNIPAAm polymer \cite{Walter_2010_2}. Therefore, the simulations can be performed with a single PNIPAAm chain. For equilibration, single PNIPAAm chains in water were simulated in the isobaric-isothermal ensemble ($NpT)$ over 1 to $5\cdot10^{7}$ timesteps. The pressure was 0.1~MPa and was controlled by the Berendsen barostat \cite{Berendsen_1984}, the temperature was controlled by the velocity rescaling thermostat \cite{Bussi_2007} and the timestep was 1~fs for all simulations. Newton's equations of motion were numerically solved with the leap frog integrator \cite{Hockney_1974}. For the long-range electrostatic interactions, particle mesh Ewald \cite{Essmann_1995} with a grid spacing of 1.2~\AA~and an interpolation order of four was used. A cutoff radius of $r_{\mathrm{c}}=15$~\AA~was assumed for all interactions. After equilibration, $2$ to $6\cdot10^{7}$ production time steps were carried out with constant simulation parameters. Note that the production steps include the conformation transition as well as the simulation of the equilibrium.\\
In order to analyze the results, the radius of gyration $R_{g}$ was calculated
\begin{equation}
  R_{g}=\left(\frac{\Sigma_{i}||\textbf{r}_{i}||^{2}m_{i}}{\Sigma_{i}m_{i}}\right)^{1/2},
\end{equation}
which characterizes the degree of stretching of the single chain, where $m_{i}$ is the mass of site $i$ and $||\textbf{r}_{i}||$ is the norm of the vector from site $i$ to the center of mass of the single chain. The radius of gyration in equilibrium was calculated as the arithmetic mean over the last $5\cdot10^{6}$ time steps of the simulation together with its standard deviation.

\subsection{Results and discussion}

\begin{figure}
  \centering
  \includegraphics[width=0.75\textwidth]{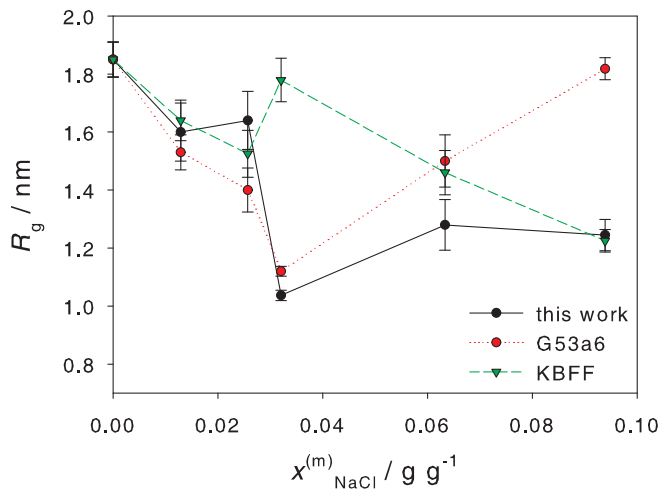}
  \caption{Radius of gyration $R_{g}$ of a PNIPAAm chain of 30 monomers in NaCl solutions of about 14,000 solvent molecules in equilibrium as a function of the NaCl concentration $x^{\mathrm{(m)}}_{\mathrm{NaCl}}$ for the different NaCl models at a temperature of 25~\ensuremath{{^\circ}}C. The error bars indicate the standard deviation.}
  \label{NaCl}
\end{figure}

For the temperature of 25~\ensuremath{{^\circ}}C, the hydrogel is swollen in pure water and collapsed above the NaCl concentration $x^{\mathrm{(m)}}_{\mathrm{NaCl}}$ of about 0.03~g$\cdot$g$^{-1}$ \cite{Zhang_2005}. Therefore, the single chain should be stretched in pure water and low electrolyte concentrations and collapsed at electrolyte concentrations above 0.03~g$\cdot$g$^{-1}$. Figure \ref{NaCl} shows the simulation results as the radius of gyration over the NaCl concentration. The NaCl model from this work and G53A6 are able to predict the $\Theta$-solvent of PNIPAAm at a NaCl concentration of 0.03~g$\cdot$g$^{-1}$. The NaCl model KBFF does not yield the fully collapsed conformation. The models, that show the $\Theta$-solvent at 0.03~g$\cdot$g$^{-1}$ also show a more or less stretched single chain at electrolyte concentrations above this point. The model with the best results here is the one from this work, for it only yields a slightly stretched conformation of the chain at electrolyte concentrations above the $\Theta$-solvent.
\\
It is therefore clearly possible to obtain qualitative predictions for the swelling of PNIPAAm hydrogels in electrolyte solutions of NaCl by molecular simulation of a single chain. For the $\Theta$-solvent in NaCl solutions, it was even possible to quantitatively reproduce experimental data. First studies on the volume transition of PNIPAAm hydrogels in electrolyte solutions of sodium sulfate Na$_{2}$SO$_{4}$ were currently carried out (results not shown here). By comparing the results of the two electrolyte solutions NaCl and Na$_{2}$SO$_{4}$ in water, it is also possible to determine the effect of the Hofmeister series on the solubility of PNIPAAm in aqueous electrolyte solutions. The Hofmeister series leads to a $\Theta$-solvent of PNIPAAm in Na$_{2}$SO$_{4}$ solutions at a lower concentration than in NaCl solutions \cite{Zhang_2005}. This correlation could be reproduced by the molecular simulations. In summary, this is an unexpectedly favorable agreement between predictions by molecular simulation and experimental data, especially when considering that the force fields were not adjusted to any such data.\\

\subsection{Computational demands}

All simulations presented in Section \ref{sec:4} were carried out with the MPI based molecular simulation program GROMACS. The parallelization of the molecular dynamics part of GROMACS is based on the eighth shell domain decomposition method \cite{Hess_2008}. With GROMACS, typical simulation runs to determine the radius of gyration in equilibrium employ 128 CPUs running for 24--72 hours. For these simulations very large systems must be considered comprising typically about 58~800 interaction sites. For these simulations a maximum memory of 284~MB and a maximum virtual memory of 739~MB was used.

\section{Conclusion}

This work covers the development of ion force fields for describing the thermodynamic properties of electrolyte solutions and the applications of these force fields for predicting the volume transition of hydrogels by atomistic molecular simulations with explicit solvent models\\
The present work proves that alkali and halogen ions can be reliably modeled by a LJ sphere with a superimposed charge located at the center of mass. The developed force fields allow predicting structural properties like the radial distribution function and the hydration number as well as thermodynamic properties like the density.\\
The self-diffusion coefficient of pure water was determined using various well known molecular models of water. The agreement with experimental data is often poor. Furthermore, ion self-diffusion coefficients were determined in simulations using these different water models. No correlation was observed between the accuracy of the self-diffusion coefficients of pure water and the accuracy of the self-diffusion coefficients of the ions. Further research effort in developing a new water model is needed to gain accurate preditions of transport properties in aqueous electrolyte systems. In addition, the study shows that the self-diffusion coefficient of ions in aqueous solutions is almost independent of the LJ energy parameter $\epsilon$ of the ion.\\
With the developed electrolyte models, it was possible to predict the volume transition of hydrogels in electrolyte solutions qualitatively and in some cases even quantitatively. The results also reproduce the effect of the Hofmeister series on the swelling of hydrogels.

\begin{acknowledgement}

The computer simulations were performed on the supercomputer HP XC4000 at the Steinbuch Centre for Computing in Karlsruhe (Germany) under the grant LAMO. This work was carried out under the auspices of the Boltzmann-Zuse Society for Computational Molecular Engineering (BZS).

\end{acknowledgement}


\end{document}